\documentclass[preprint,aps]{revtex4}
\usepackage{amssymb,amsmath}
\usepackage{graphicx,float}

\begin{document}
\title{ \Large \bf Transport Processes on Homogeneous Planar Graphs with Scale-Free Loops }
\author{\large  \flushleft 
Milovan \v Suvakov and Bosiljka Tadi\'c} 

\affiliation{ 
$^1$Department  for Theoretical Physics, Jo\v{z}ef Stefan Institute; 
P.O. Box 3000; SI-1001 Ljubljana; Slovenia, 
\\ \hspace{1cm}
} 

\begin{abstract}

We consider the role of network geometry in two types of diffusion processes:
transport of constant-density  information packets 
with queuing on nodes, and constant voltage-driven tunneling of electrons.
The underlying network is a homogeneous graph with scale-free distribution of loops, which is constrained to a planar geometry and fixed node connectivity $k=3$. 
We determine properties of noise, flow and return-times statistics for both 
 processes on this graph and relate the observed differences to the 
microscopic process details. Our main findings are:
(i) Through the local interaction between  packets queuing at the same node,
 long-range correlations build up  in traffic streams,
which are practically absent in the case of electron transport; 
(ii) Noise fluctuations in the number of packets and in the number of 
tunnelings  recorded at each node appear to obey the scaling laws in two 
distinct  universality classes; 
(iii) The topological inhomogeneity of betweenness  
plays the key role in the occurrence of 
broad distributions of return times
and in the dynamic flow. The  maximum-flow spanning trees are characteristic for each process type.
\end{abstract}

\maketitle
\section{Introduction}
In recent years an intensive research on networks structure and dynamics \cite{recent_review} was conducted, motivated both with theoretical reasons and 
practical applications to improve network's performance, for instance
 in the communication networks. It has been recognized that network structure
may influence the course of the dynamic processes on it through the 
{\it topological constraints} fixed by network complex patterns of links 
and higher 
topological sub-structures. Certain types of processes run better on networks with higher structural complexity \cite{we_review,TT,TTR}, whereas other processes,
like synchronization \cite{synchronization}, 
 are  more efficiently accomplished  on homogeneous structures. In this respect, a range of structure--function matching
possibilities opens, depending on the details of the dynamics \cite{BCN_optimization,Hurtado,KK}. 
 
In the case of transport of information packets
two markedly different structures were found \cite{BCN_optimization} corresponding to minimized travel time of packets: a scale-free structure at low traffic density, and a homogeneous structure at high traffic density. The highly clustered scale-free graph was also found to support efficient  free-flow traffic with the local navigation rules compatible with the
``information horizon'' of the graph \cite{TT,TTR,TT_05}. The emergent
efficiency of transport is 
related to the central role of the hub and its associated super-structure 
\cite{TTR}.
However, the structure is vulnerable for jamming when the creation rate exceeds a  (large) critical value \cite{TT,TTR}. In this regime, a 
homogeneous network
with distributed node activity is expected to perform better  
\cite{BCN_optimization}.
Apart from the limitation to {\it free graphs} (i.e., graphs in the
infinite space dimension) and relatively low transparency of the simulated annealing procedure at large networks, the results of Ref.\ \cite{BCN_optimization} 
revealed main structure--traffic interdependences and 
traced routes towards improved efficiency of transport processes
on complex networks. The following questions remain open: 
\begin{itemize}
\item{} What topological property  on a small scale of the underlying network  
plays the key role in given dynamics? 
\item{} How these globally optimal topologies  perform 
in the case of traffic with constraints of restricted geometries?
\end{itemize}
Here  we address these questions using   numerical simulations to study comparatively  two types of
transport processes on a {\it homogeneous planar graph with scale-free distribution of loops}. In particular, we  study:  
(1) Traffic of information packets 
between specified pairs of nodes on the graph at constant packet density 
 (fixed number of moving packets) $\rho \gg 1$; 
 and (2) Transport of electrons via tunneling between nodes, which are driven by the constant voltage difference $V=const$ on electrodes. 
In both cases the transported packet (electron) 
has a specified source and sink as a pair of nodes on the graph. 
The packets are navigated locally towards their destination (sink) node, whereas the electrons are driven globally by the voltage profile. 
Another important difference in these processes is that the information packets 
interact by  making queues on a node. In contrast, the number of  electrons 
on a node is practically  unrestricted and order of their processing is random. 
We determine quantitative characteristics of these two processes on the same network (fixed by its adjacency matrix) and 
attempt to relate the statistical properties of transport to the microscopic 
process rules. 
In contrast to the above mentioned networks in {\it infinite dimensional
 space}, we consider here a graph that is constrained to a {\it planar geometry}. We also 
fix the number of links per node to strictly $k=3$, in order to minimize the
effects of node connectivity. 
Our main findings are that in transport on planar graphs the key topological feature is the betweenness centrality of nodes. Different processes utilize the underlying structure in different ways leading to statistically different outcomes.
The observed traffic properties  can be further related to certain features of the
 microscopic dynamics, specifically to queuing and driving details.

In Section II we present the graph structure and its main topological characteristics \cite{MSBT_iccs06}. Section III defines the transport processes and their
numerical implementation on the graph. Main features of the traffic noise
are considered in section IV, and relation between the topological 
and dynamical centrality measures are studied in Section V. Section VI contains a short summary and the discussion of the results.

\section{Structure}

Planar graphs are mathematical objects that can be embedded into an Euclidean plane. They obey the following topological
 constraints \cite{BB_book}: A graph is planar {\it iff it does not contain a subdivision of $K_5$ (5-clique) and $K_{3,3}$ structures},
and the Euler relation 
 $ N_p+N=E+1$ between   the number of polygons $N_p$,   nodes $N$, 
and  links $E$ is satisfied. 

We grow a planar graph by {\it cell aggregation} procedure, which we introduced earlier \cite{MSBT_iccs06}. Here we observe a strict constraint of node connectivity $k=3$ for all nodes enclosed in the graph interior.
The cells are loops of nodes with varying lengths
$\ell$ 
driven from a power-law distribution $P(\ell ) \sim \ell ^{-\tau}$. 
The aggregation is starting from an initial loop and nesting additional loops
along the boundary of the graph. In order to observe the constraint
of maximum connectivity $k=3$, a nesting place along the graph boundary
is searched and its length $n_d$ is identified as the distance between two successive nodes with connectivity $k=2$.
Then the nesting is performed 
with a probability
$p\sim \exp {(-\nu n_s)}$, where $n_s$ represents the excess number of nodes,
defined 
as the difference between current loop length $\ell $ and the length of the selected nesting string $n_d$.
The number of nodes that are successfully nested are identified with the
existing  $n_d$ nodes along the nesting site on the graph boundary. 
In this way the parameter $\nu$ plays the role of the chemical potential for addition of nodes. In general, when $\nu $ is small, the probability of adding a large number of new nodes $n_s$ attached to a short nesting site is larger, resulting in a more branched graph structures, compared with more homogeneously space-filling loops obtained for  large $\nu$ values \cite{MSBT_iccs06}. 
An example of the structure for large $\nu $ and $\tau =2.2$ is shown in Fig.\ 1 (left).

\begin{figure}[htb]
\begin{center}
\begin{tabular}{cc} 
\resizebox{19pc}{!}{\includegraphics{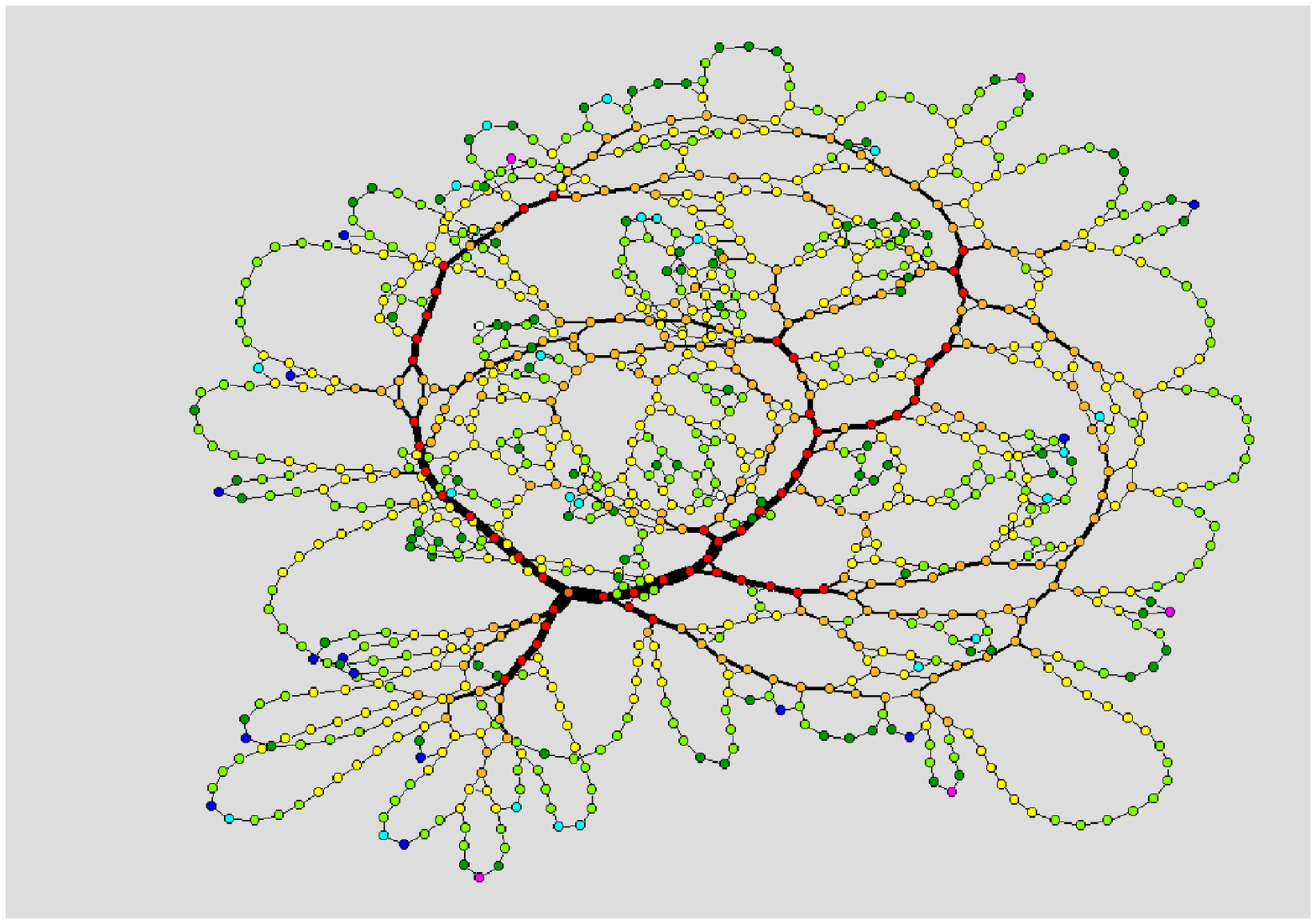}} &
\resizebox{19pc}{!}{\includegraphics{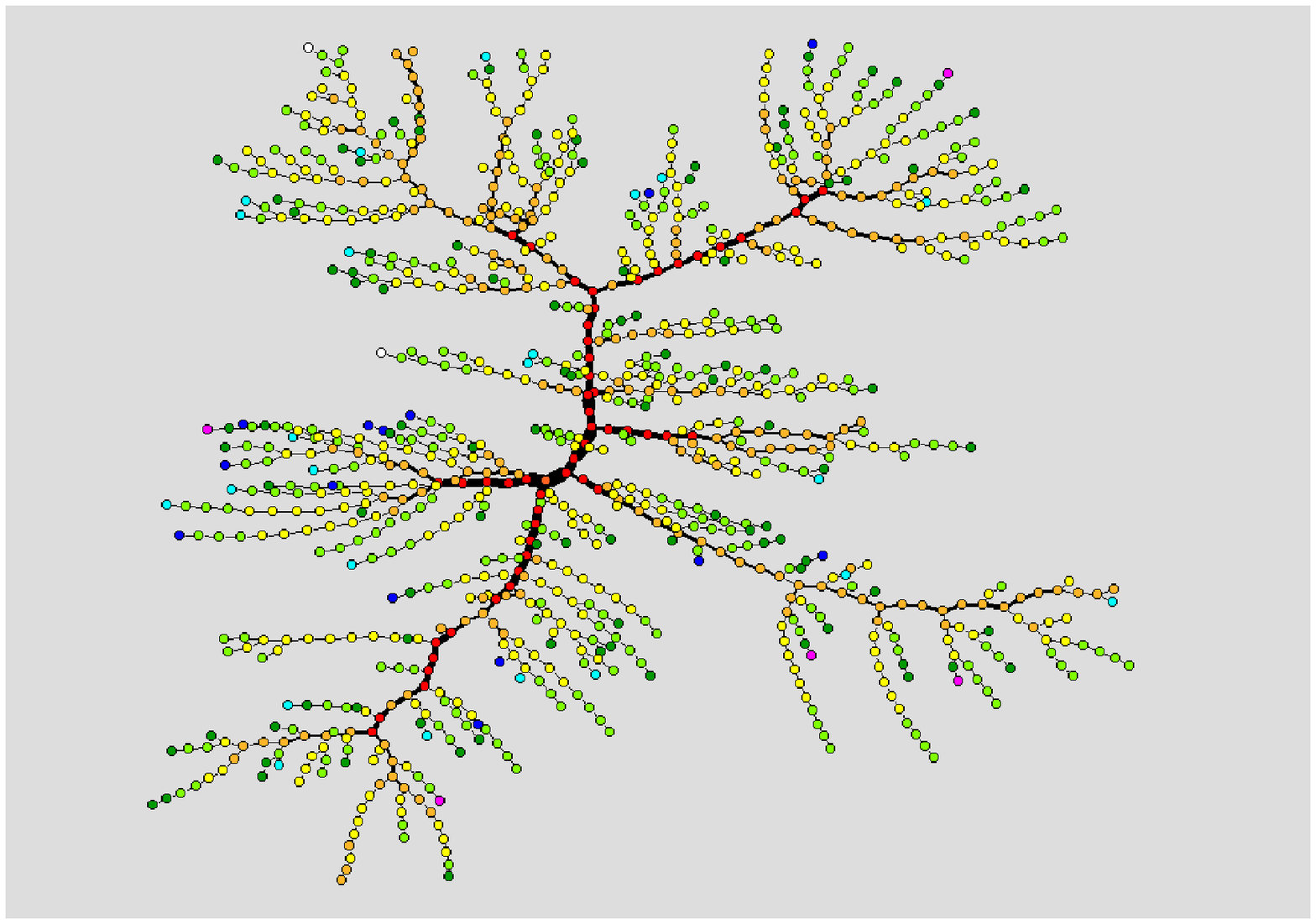}} \\
\end{tabular}
\end{center}
\caption{Emergent planar graph with scale-free distribution of loop sizes and nesting potential  $\nu =5$ (left) and its maximum-betweenness-spanning 
tree (right). Line widths indicate relative topological betweenness of links. Different colors are coding the topological centrality of nodes. }
\label{figgraphs}
\end{figure} 

It should be stressed that, due to the planar constraints the emergent graph structure is {\it not of a small-world type}. In addition, the graph is strictly homogeneous and clustered (probability of a short triangle loop $\ell =3$
is relatively high in the power-law distribution of loops). 
A remarkable feature
of this type of graphs is its  inhomogeneity in  the betweenness centrality 
for both nodes and links. In Fig.\ 1 the centrality of links is marked
 on the graph with lines of different widths. The topological {\it maximum betweenness spanning tree}, in which each node is linked to the rest of the nodes on the graph via its {\it largest  betweenness link}, is also shown in Fig.\ 1 (right). Due to graph homogeneous connectivity, the maximum betweenness
nodes lie on the maximum betweenness links (see the color-coded figure).

\section{Dynamics}

We consider two types of dynamic processes on the graph shown in Fig.\ 1 (left): (i) Autonomously driven constant-density transport of information packets (IP)
 and  (ii) Constant-voltage- driven tunneling current (Q). 
Next we describe the two processes in detail and then determine their   statistical properties,
in particular features of the traffic  noise and dynamical flow. We then discuss the traffic features relative to the 
topological properties  of the graph.  
The graph (Fig.\ 1) contains $N=1003$ nodes connected according to its adjacency matrix and  is identical in both processes.

\subsection{Information Flow at Constant Density}

Transport of the information packets on the graph takes part between an in advance specified pair of nodes---origin and destination node of a packet, which are selected randomly.
Motion of a packet on the  network is implemented as a 
{\it guided random walk} from the origin of the packet towards its destination
\cite{TT,TTR,TT_05,BT_ccp03}. At each node the packets are navigated 
through the graph using the local 
{\it nnn}-search rule, where  two depth levels around each node 
(sometimes called information horizon 2) are searched for the
packet destination address \cite{TT,TTR,TT_05,BT_ccp03}. 
The rule is supplemented by random diffusion
when the search is unsuccessful. The whole network is updated in parallel.
The packets are removed when they arrive at
their destinations. Here we implement the traffic for a
{\it fixed number of moving packets}. We start with a given number 
$\rho$ of packets. The arrived and 
removed packets are replaced in the next time step
by creating the same number new ones at randomly chosen nodes and assigned new destinations. In the limit $\rho =1$
this corresponds to the  sequential guided random walk problem. 
For  $\rho > 1$ packets interact with each other by making queues at nodes along their paths. The queuing of packets is a qualitatively new feature of {\it dense traffic} on networks. The packet density is controlled with the number of moving packets $\rho$ on the graph with a fixed number of nodes $N$.
We assume finite queue lengths $H=1000$, and a LIFO (last-in-first-out) queuing rule \cite{TTR}.  Details of the numerical implementation can be found in Ref.\ \cite{BT_ccp03}.

\subsection{Voltage-Driven Electron Tunneling}

{\it Description:} In this dynamic process the graph represent system of nano particles (nodes) \cite{Philip_SA,Philip_REP} with equal inter-dot capacitances over 
links, $C$, and equal capacitances between dot and gate, $C_g$.
We assume conditions that lead to Coulomb-blockade transport \cite{nanoarrays_book} with equal tunneling resistance $R$.
Electrodes are associated to two subsets of the graph boundary  nodes. 
Each electrode length and distance between them is taken as one quarter of  the graph boundary length. Total energy of this system can be written as
\cite{nanoarrays_book,PRL-93,CB}
\begin{equation}\label{energy}
E=\frac{1}{2} \mathbf{Q}^\dag M^{-1} \mathbf{Q} +  \mathbf{Q} \cdot V^{ext}  + 
Q_{\nu} \Phi^{\nu},
\end{equation} 
\begin{equation}\label{vext}
V^{ext}=M^{-1} \mathbf{C}_\nu \Phi^{\nu},
\end{equation} 
where $M$ is the capacitance matrix, $\mathbf{Q}$ is the vector of charges on dots, $\Phi^{\nu}$ ---potential of an electrode, $\nu \in \lbrace +,-,gate \rbrace$, and $\mathbf{C}_\nu$ is the vector of capacitances between dots and electrode $\nu$. The diagonal elements of $\mathbf{M}$ are the sum of all capacitances associated with a dot and off-diagonal elements are the negative interdot capacitances: $M_{ij}=-C_{ij}$, $i \neq j$ and $M_{ii}=\sum_j C_{ij}+\sum_\nu C_{\nu,i}$, where $C_{ij}$ is the adjacency  matrix of the graph.

{\it Dynamics:} The single electron tunneling rate through a junction (link) 
$ij$ is calculated:
\begin{equation}
\Gamma_{ij} = - \frac{\Delta E_{ij} / e^2 R}{1-\exp({\Delta E_{ij}}/{k_BT})},
\label{tunneling_rate}
\end{equation} 
where $R$ is the quantum tunneling resistance, $\Delta E_{ij}$ is the energy change associated with a tunneling process along the  $i\to j$ link,  and
$k_BT$ is the thermal energy. The  rate $\Gamma_{ij}$ then determines the 
probability density function of the tunneling time $t$  
by  $p_{ij}(t)=\Gamma_{ij} \exp(-\Gamma_{ij} t)$. 

{\it Implementation:} System is initiated as a network of dots without 
charges and a voltage difference $V$ between the electrodes is imposed. 
In each step we compute 
rates $\Gamma _{ij}$  and determine respective tunneling times
from the distribution $p_{ij}(t)$ for all junctions $\{ij\}$.
 We process a single electron tunneling through the junction that has the 
shortest tunneling time. After each step we sample time, number of charges on all dots, number of charges that arrived to the zero-voltage electrode (current). We also 
keep track of number of tunnelings that occurred along  each  link and on 
each node.

{\it Parameters and optimization:} In all calculations the potential of the gate and on 
one electrode is kept to zero and  constant potential $V$ on the other 
electrode.  The system was on zero temperature.
The elements of the inverse capacitance matrix $M^{-1}$ fall off exponentially with a screening length, which is proportional to $C/C_g$. We assume
$C/C_g \ll 1$, which permits  
computational acceleration by using only nearst-neighbour elements of 
$M^{-1}$ 
for  calculation of the energy change in Eqs.\ (\ref{energy}-\ref{tunneling_rate}). 
For the same reason, the potential 
 $V^{ext}$ (\ref{vext}) can be considered as a constant 
 within the shortest computed time window $\delta t \sim 1/\Gamma_{max}$.

{\it Non-linearity and statistics:} Driven by the external potential, the 
charges are entering  the system from the high-voltage  electrode. For small voltage on that electrode, the charges  can screen-off the external potential. In that case, after some transient time, the system become static and there is no current. 
For voltage larger than a threshold voltage $V_T$ the screening does not occur
 and after some  transient time a stationary distribution of charges over dots
sets-in  with a constant current.
In general, current-voltage dependence is  non-linear \cite{PRL-93,we-future}
$I \sim (V-V_T)^{\zeta}$ for $V > V_T$. For this network type we find 
  $\zeta \approx 3$ \cite{we-future}. In this paper we consider transport 
properties for a {\it fixed voltage} $V=10V_T$ in the non-linear regime.
The data are collected after the transient time when a  constant current occurs.

\section{Noise Properties in the Stationary Flow Regime}

In the case of electron tunneling with voltage difference $V$ between the
electrodes, the distribution of electrons per node in the stationary state 
is compatible with the 
actual value of the voltage at that node. An electron  moves along a 
link towards the zero-voltage electrode when the tunneling condition $\Gamma_{ij} > 0$, with $\Gamma_{ij}$ in Eq.\  
(\ref{tunneling_rate}), is  satisfied. Thus the number of tunnelings on a node 
fluctuates 
in time around an average value which is determined by the voltage profile.
We consider  the charge fluctuations $Q_i(t)$ at a node
 $i$ as a time-series taken over  discrete time points $t/<\Gamma_{max}^{-1}>=1,2, \cdots $.

In the case of information packet transport the stationarity is guaranteed by the implementation of the constant density, here $\rho = 100$ packets.
(In other driving modes  the stationary traffic occurs for posting rates below a certain critical value for  each network \cite{TT,TTR}.)
For a fixed node $i$ on the network, the number of
packets arriving to that node $h_i(t)$ fluctuates in time steps $t$, depending on the activity of its neighbour nodes. Each node tries to process top packet on it 
towards packet's actual destination. Thus a macroscopic current is absent.

 {\it Power Spectra.} The power spectrum of the corresponding time series is shown in Fig.\ 2a (bottom line), with  the form $S(f)\sim f^{-\phi}$, $\phi \approx 0.9$, within error bars,indicating a long-range correlations  for large frequencies. In contrast, the corresponding power-spectrum of the time series
collected at the zero-voltage electrode in the tunneling process, which is 
given  in Fig.\ 2b (lower curve), indicates  a white noise.
Another time series comprises of the fluctuating number of
nodes which are simultaneously processing a packet (electron), $n(t)$.
Its  power spectrum is also shown in Fig.\ 2a and b (top curves) for both packet and electron processing, respectively. These  series appears to be weakly correlated on the planar graph. 
Since the network structure is identical in both processes, the observed difference in the noise correlations can be attributed to the interaction mechanisms and driving conditions. In particular, the processed packets interact locally via queuing at a node, whereas number of electrons at each node is determined by global
energy minimum conditions, given by the Hamiltonian in Eq.\ (\ref{energy}). 
\begin{figure}[htb]
\begin{center}
\begin{tabular}{cc}
\resizebox{19pc}{!}{\includegraphics{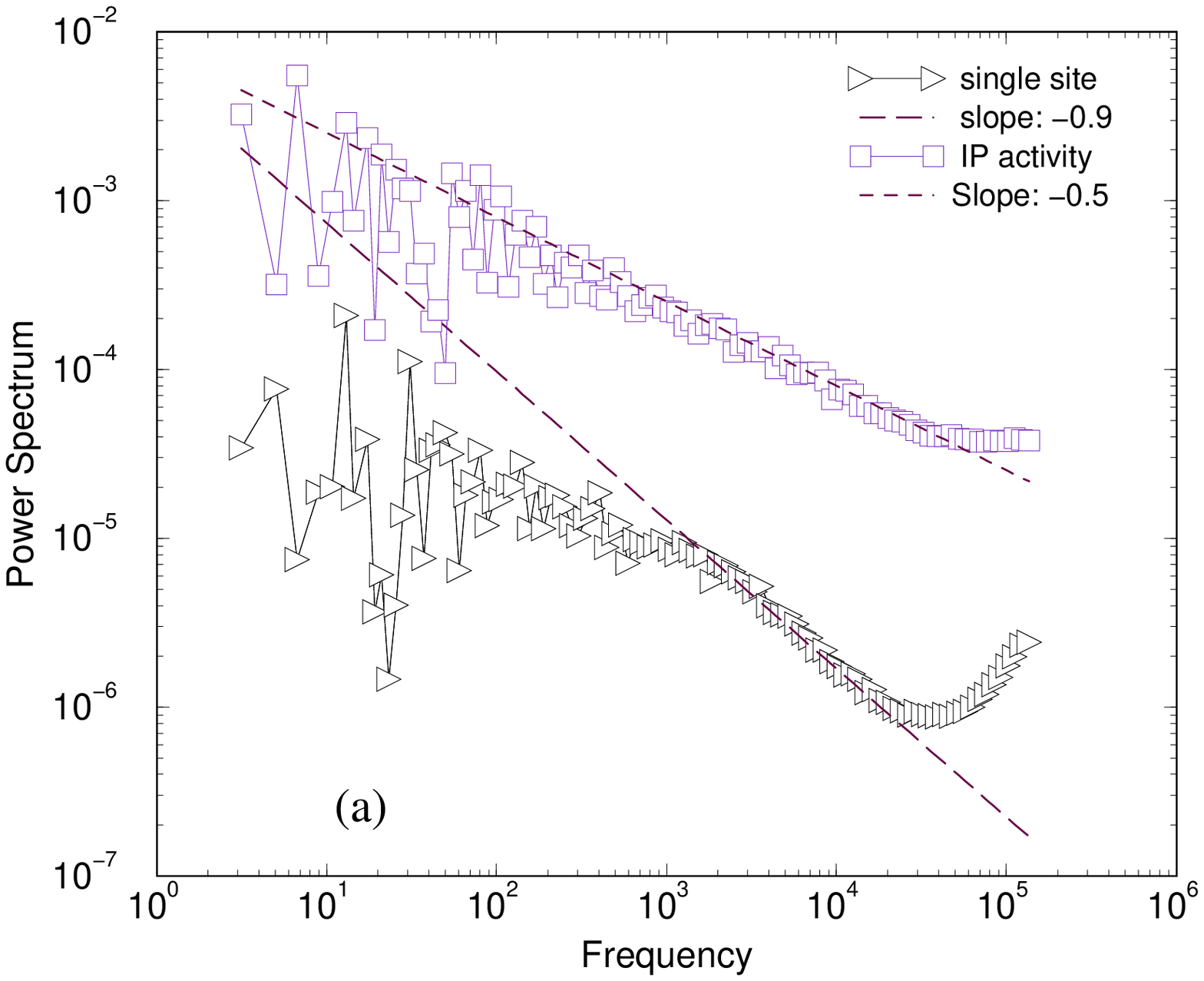}}&
\resizebox{19pc}{!}{\includegraphics{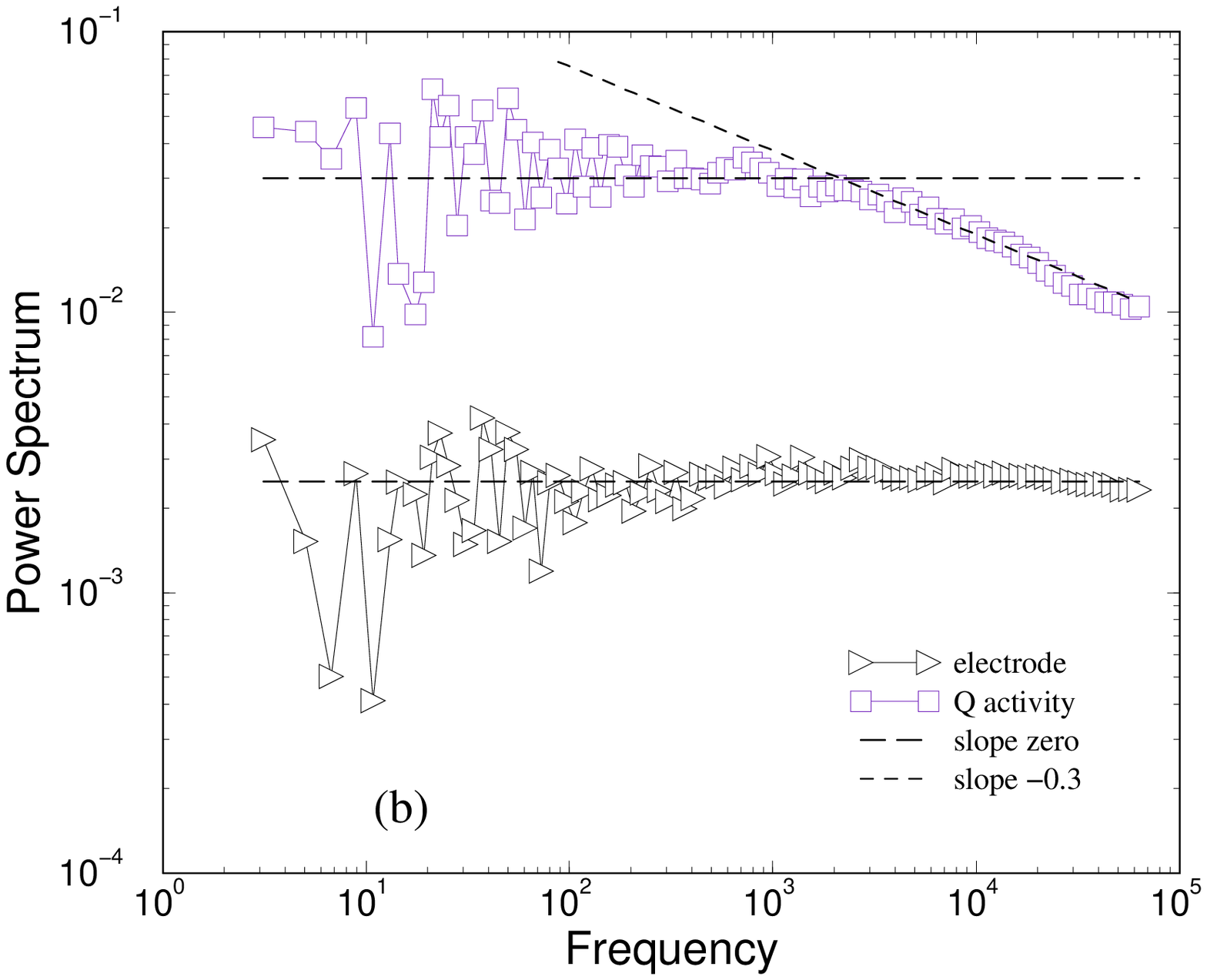}}
\end{tabular}
\end{center}
\caption{Power spectra of the recorded time series of activity at a single node (lower curves) and of a number of active nodes for transport of information packets 
(a) and tunneling electrons (b).}
\label{fignoisecorrelations}
\end{figure} 

\begin{figure}[htb]
\begin{center}
\begin{tabular}{cc}
\resizebox{19pc}{!}{\includegraphics{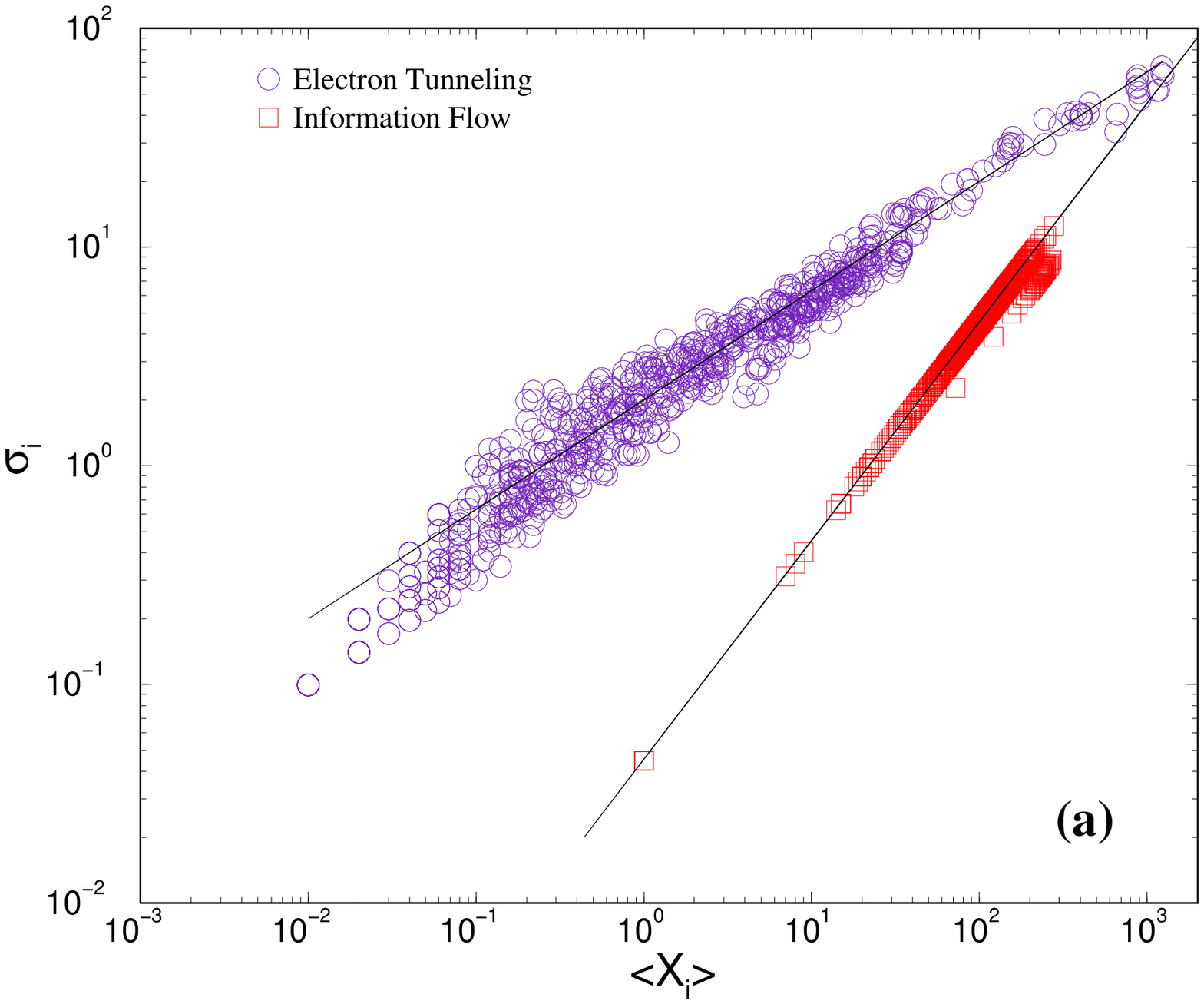}}&
\resizebox{19pc}{!}{\includegraphics{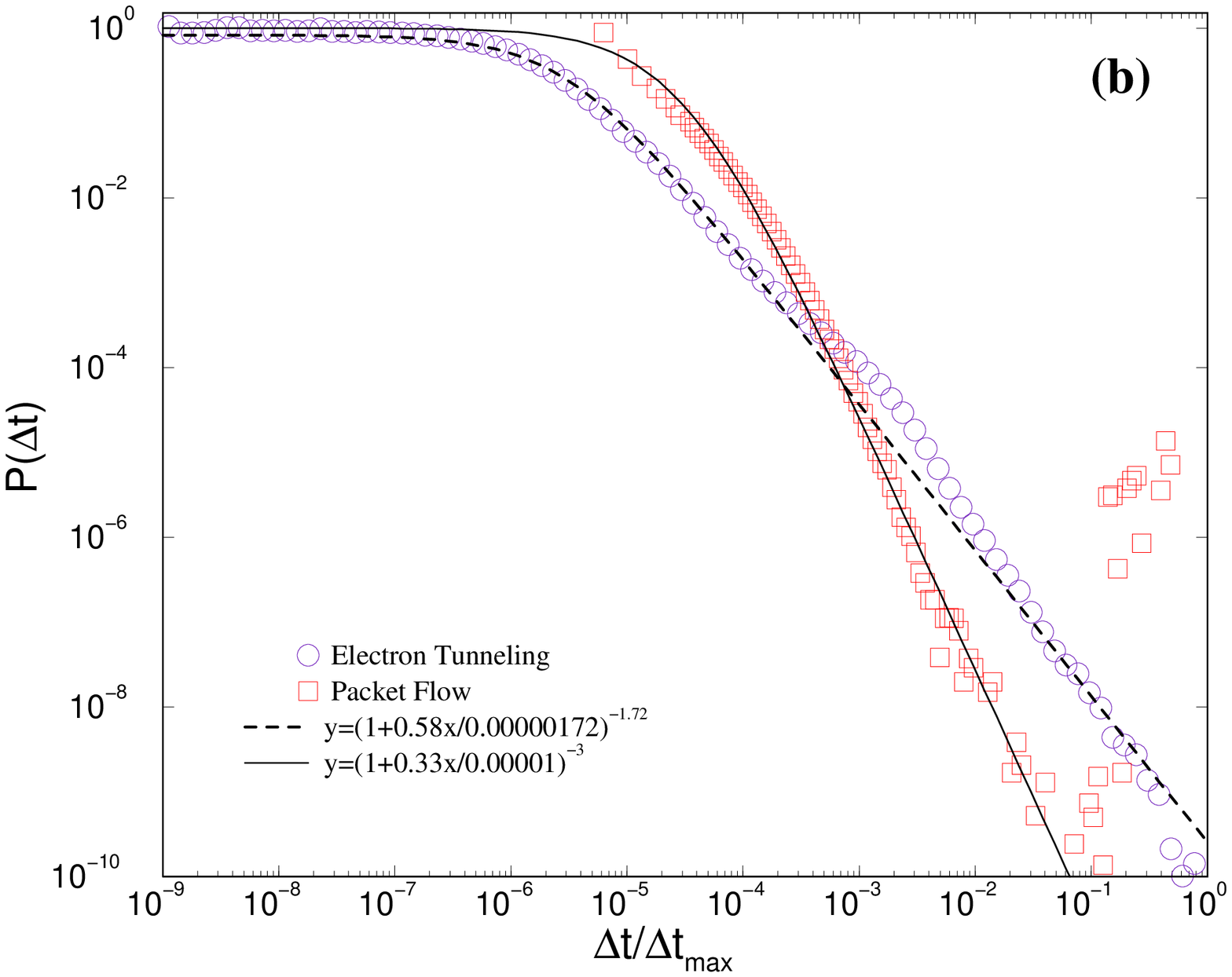}} \\ 
\end{tabular}
\end{center}
\caption{(a) Standard deviation $\sigma _i$ against average $<X_i>$ of the number of processing events at a node $i$ within time window $T_{win}=1000$ steps for the information packets  and electron tunneling  at cellular network in Fig.\ 1. Full lines indicate slopes $\mu = 1/2$ and $\mu =1$. (b) Distributions of return-time intervals for the packet and electron transport. Fits are according to Eq.\ (\ref{q-exp}).}
\label{fig_hs-rett}
\end{figure}
 {\it Multi-channel Noise.} Monitoring the number of packets processed by each node within a fixed time window of $T_{win}=1000$ time steps, we make a set of time series $\{h_i(t_w)\}$ for each node
$i=1,2, \cdots  N$, where $t_w=1,2, \cdots $ enumerates consecutive time windows. The  computed standard deviation $\sigma _i$ of the time series $h_i(t_w)$ for each node $i=1,2, \cdots N$ separately is shown against its  average over all time windows 
$<X_i>\equiv <h_i(t_w)>$ in Fig.\ \ref{fig_hs-rett} (a). It appears that  the scaling law \cite{scaling_hsigma}
\begin{equation}
\sigma _i \sim <X_i>^\mu \ , \label{h-sima}
\end{equation}
holds for  all nodes, with the exponent $\mu =1$.
With  analogous analysis  of the set of time series $\{Q_i(t_w)\}$ and $<X_i> \equiv <Q_i(t_w)>$  
for the electron tunneling process at all nodes, we find that the 
scaling law (\ref{h-sima}) also holds,
however, with clearly different  exponent $\mu = 1/2$, as shown in Fig.\ \ref{fig_hs-rett} (a).

{\it Return-Times Statistics.} In order to further substantiate the difference between the two transport processes, we determine the return-time statistics, which demonstrates how often the  process occurs at a given node. 
The return time $\Delta t _i$ is defined as time interval between two 
consecutive events at a fixed node $i$. 
The distributions of  return times appear to have  power-law tails, that can be fitted with the $q$-exponential form \cite{Tsallis}
\begin{equation}
P(\Delta t) = B_q\left[1 -(1-q)\Delta t /\Delta t_0\right]^{1/(1-q)} \ 
\label{q-exp}
\end{equation}
with $q=1.58$ for the electron tunneling, and $q=1.33$ for the packet transport.  The results are shown in Fig.\ \ref{fig_hs-rett} (b) for a reduced variable $\Delta t/\Delta t_{max}$, with $\Delta t_{max}$ is maximum observed interval for given
 process. This statistics suggests that  
 longer waiting times and thus stronger correlations  (larger $q>1$ values) occur more often in the electron tunneling. However, the corresponding  noise fluctuations at individual nodes are compatible with $\mu =1/2$, contrary to 
the arguments  given in Ref. \cite{scaling_hsigma}. 

In the packet processing, the local search is ineffective on the homogeneous
(non-small world) network, and packets perform mostly random walk involving a large number of nodes (number of active nodes fluctuates around 94 for the density $\rho=100$). However,  the packets interact due to queuing  practically at all nodes along the path. The number of arriving packets at each node is limited to maximum three (number of links), whereas one packet is forwarded per time step. These transport constraints result in
the noise fluctuations with the exponent $\mu =1$  on our planar graph. 

\section{Role of Betweenness Centrality in Transport Processes}

Contrary to the small-world graphs, in a homogeneous planar graph  
the even connectivity of nodes does not guarantee their
even betweenness centrality. 
The  betweenness centrality of nodes (links) in our network as shown in 
Fig.\ \ref{figgraphs} appeares to be  widely distributed. This topological inequality of nodes (links)
is the basis for a  different role that individual nodes (links) play in a dynamic process on that network. Additional difference developes due to microscopic details of the dynamics. Here we compare the topological betweenness centrality
with dynamical flow for the two types of processes described above.

The computed topological betweenness centrality of nodes (links) of the network is also displayed in Fig.\ \ref{figgraphs} via a color-code (exhibiting the spectrum
from red, orange, yellow, green, blue, deep-blue and purple) for descending
betweenness centrality in a suitably chosen log-binns. Note that, 
due to the constant node connectivity,   the 
betweenness of links follows a similar pattern, so that nodes of high 
betweenness are placed along the links with high betweenness 
(shown with different 
widths of links in Fig.\ \ref{figgraphs}). 
The maximum-topological-flow spanning tree (shown in Fig.\ \ref{figgraphs} right) connects pairs of nodes via the most frequent  paths on the graph. Obviously, the skeleton of  the tree 
represents the most frequent part of all paths on the graph. Nodes with largest centrality are placed along that path stretch.
\begin{figure}[htb]
\begin{center}
\begin{tabular}{cc} 
\resizebox{18pc}{!}{\includegraphics{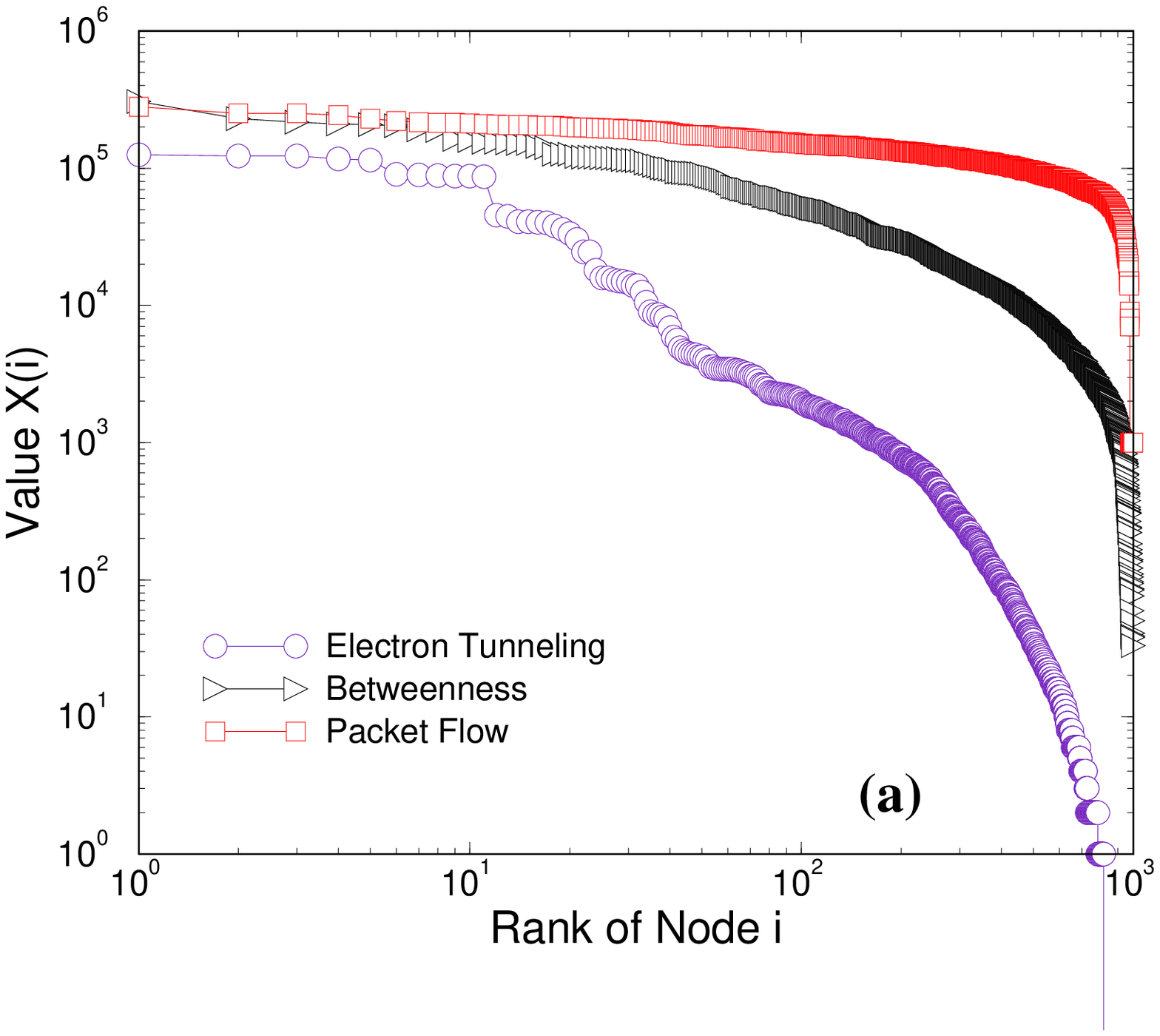}} &
\resizebox{18pc}{!}{\includegraphics{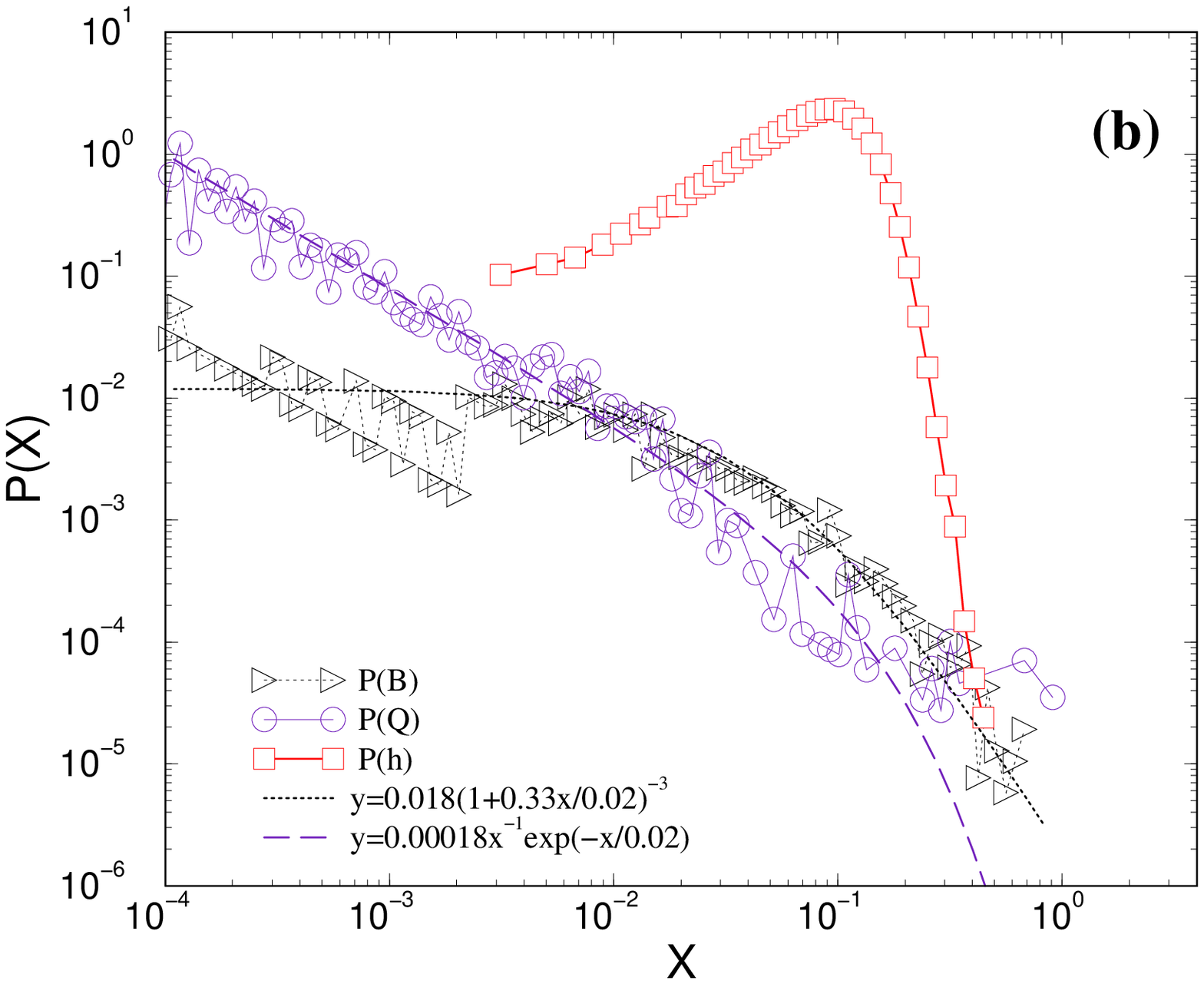}} \\
\end{tabular}
\end{center}
\caption{(a) Ranking statistics of nodes according to  their
topological betweenness centrality ($\triangleright$), dynamical  number of 
tunnelings  ($\bigcirc$) and number of processed packets ($\square$). 
(b) Distributions of topological centrality ($\triangleright$), number of 
tunnelings ($\bigcirc$), and number of processed packets ($\square$).
}
\label{fig_betw_rank}
\end{figure} 
\begin{figure}[htb]
\begin{center}
\begin{tabular}{cc} 
\resizebox{19pc}{!}{\includegraphics{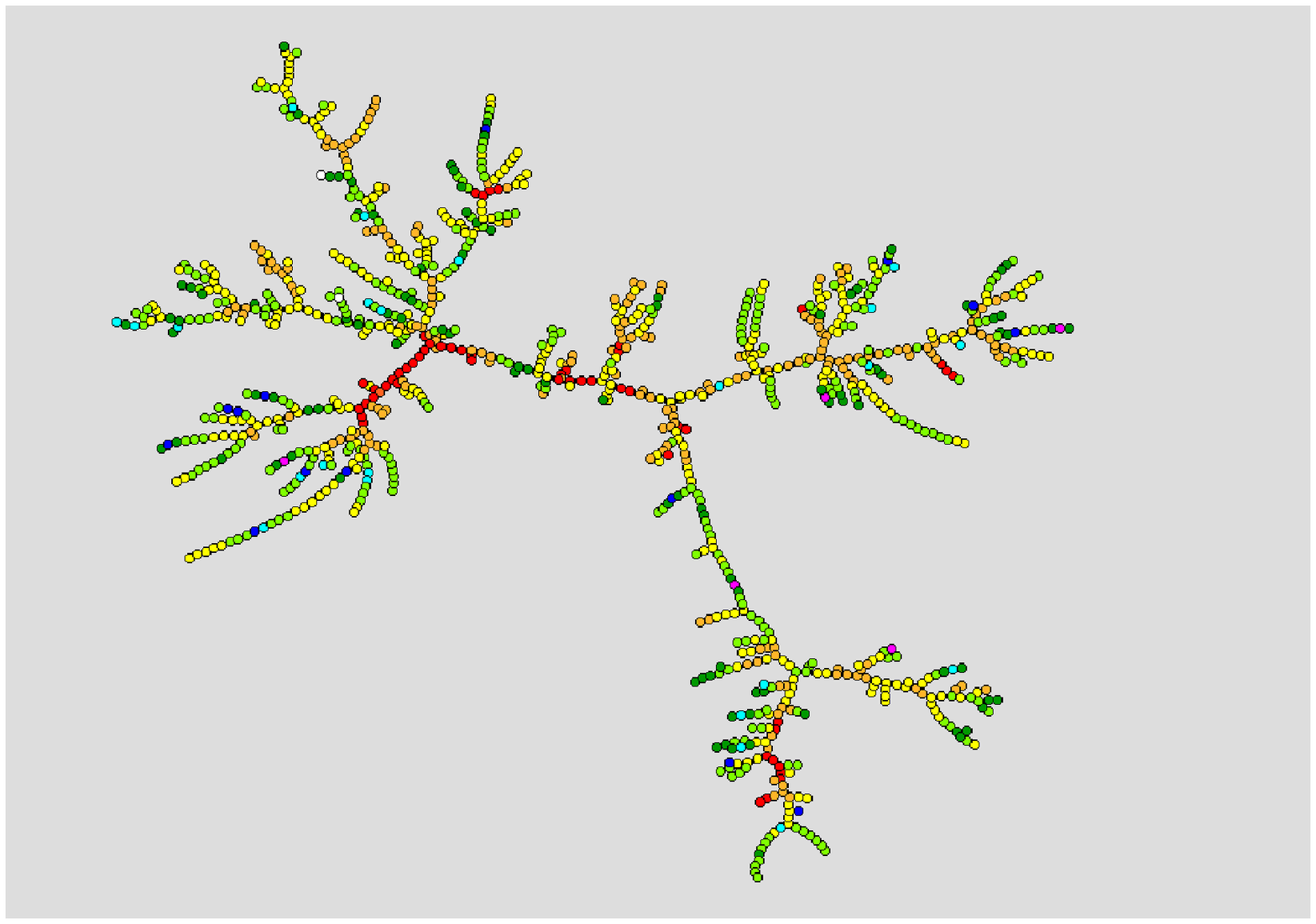}} &
\resizebox{19pc}{!}{\includegraphics{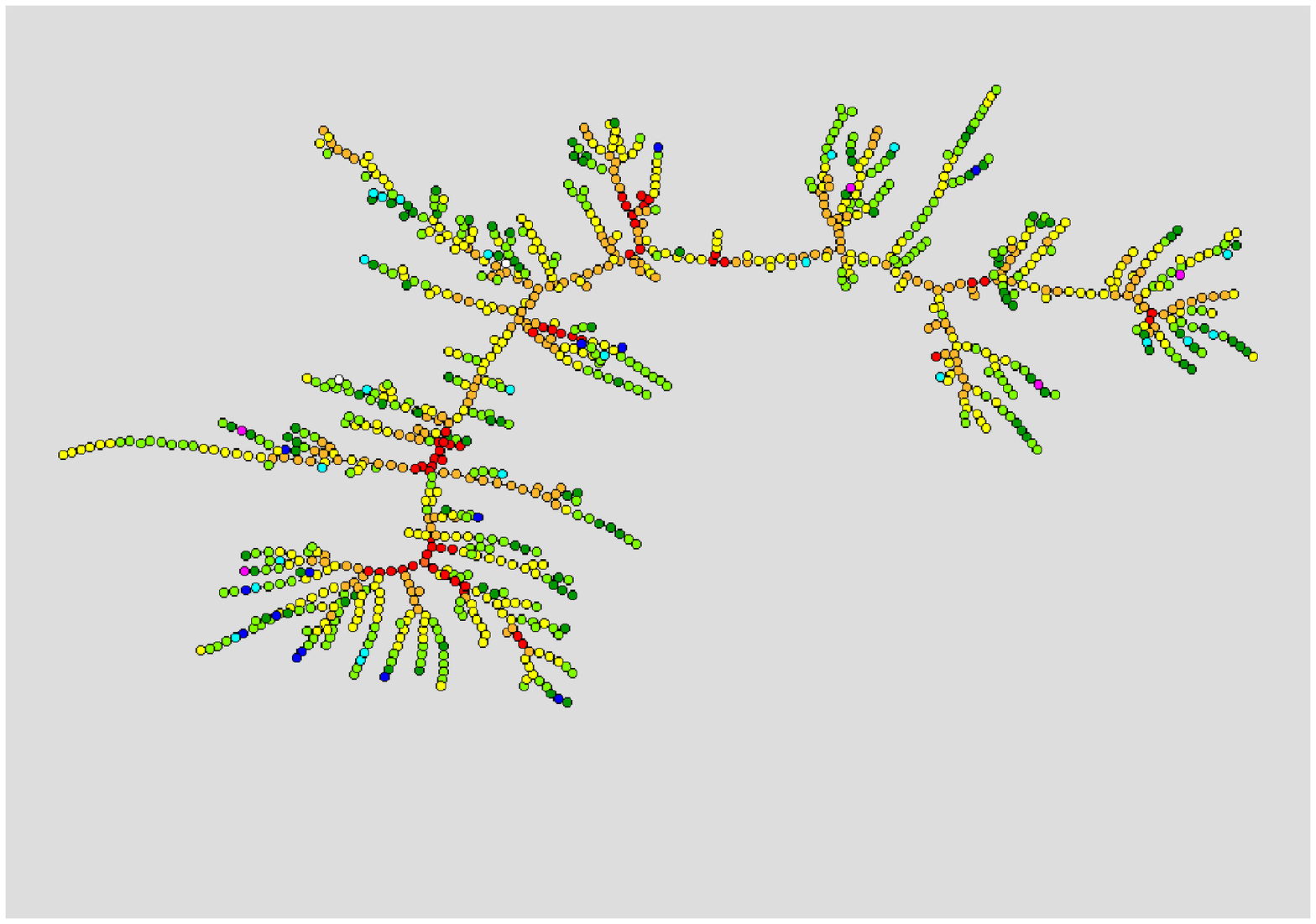}} \\
\end{tabular}
\end{center}
\caption{Information packets maximum-flow spanning tree (left)
and tunneling-current maximum-flow spanning tree (right).
Color-code represents topological betweenness of nodes, as in  Fig.\ 1.}
\label{figgraphs_flowtrees}
\end{figure} 
Additional details of the  quantitative analysis  are shown in Fig.\ 
\ref{fig_betw_rank}. 
The inhomogeneous betweenness of nodes results in a broad ranking-statistics 
curve (stretched-exponential) and a broad  distribution of betweenness
(Fig.\ \ref{fig_betw_rank}b). Compared to the topological betweenness, the dynamical betweenness
(or flow), which is defined as number of packets processed by a node,
in the case of packet transport appears to be more even in the node ranking 
statistics. The picture   is compatible with the random walk dynamics,
in which all nodes become almost equally busy in the transport process. This  
results in a narrow distribution of the dynamical
information-flow betweenness, as shown in Fig.\ \ref{fig_betw_rank}b.
In the case of electron tunneling, the profile of the ranked dynamical 
betweenness of nodes
is steeper, and consequently the distribution broader,
 compared to the topological betweenness. The reason is the directed flow
of  electrons between the electrodes, and additionally, the reduced active
part of the graph, due to the selected position of the electrodes.

The difference between these two processes is  also illustrated by their maximum-dynamic-flow spanning trees, which are shown in Fig.\ 
\ref{figgraphs_flowtrees}. Obviously, the two dynamic processes use the 
underlying graph topology in different ways.
Nevertheless, the nodes with maximum topological betweenness are  often placed along the central line on the dynamic flow trees, suggesting the key role of these nodes in both  transport processes.

\section{Conclusions}

We have studied two dynamic processes on a homogeneous planar graph with scale-free loops
in which we 
demonstrated several aspects of {\it constraints in the network dynamics}.
First, the growth of the graph is  constrained to a planar geometry and
fixed node connectivity. This results in a non-small world graph, in which topological betweenness centrality of nodes (links) appears to be inhomogeneous, making the basis for  nontrivial dynamical effects.
Second,  the graph's geometry makes the constraints to the 
dynamic processes. 

In the two types of diffusion processes---information packet flow and 
voltage-driven
electron-tunneling---which take part between pairs of nodes on the graph,
details of the dynamics result in different use of the underlying graph
topology. The relevant features at microscopic scale are, on one side, 
random-walk dynamics 
with queuing of information packets, against directed transport of  
electrons with global constraint on the number of electrons at a node,
 on the other. 
In particular, we demonstrated the emergent  quantitative difference 
in noise correlations,  universal noise fluctuations and return-time 
statistics, and in the respective dynamical maximum-flow spanning trees.
Our numerical data, for instance for the return-time distributions and  
the noise fluctuations,  appear to be well fitted with theoretical expressions
suggested in different concepts of the complex dynamical systems 
\cite{scaling_hsigma,Tsallis}.  The full understanding of the occurrence of these laws in the transport on networks remains elusive.
We hope that our numerical study may serve as a basis for  further 
theoretical research to unravel  the role of constraints in the dynamic 
processes on networks.

\acknowledgments
B.T. thanks for support from the Program P1-0044 
of the Ministry of high education, science and technology, Slovenia; 
 M.\v S. is supported from the Marie Curie Research and Training Network 
 MRTN-CT-2004-005728 project.

\end{document}